\documentclass[12pt, titlepage]{article}

\usepackage{amsmath}
\usepackage{amssymb}

\usepackage{graphicx}

\usepackage[round,numbers,sort&compress]{natbib} 
\begin{document}
\date{}

\title{Cell Growth and Size Homeostasis in Silico}

\author{Yucheng Hu\thanks{Corresponding author. E-mail: huyc@tsinghua.edu.cn}$^{,2}$ \\
Zhou Pei-yuan Center for Applied Mathematics,\\ 
Tsinghua University, Beijing, China, 100008
\and Tianqi Zhu\thanks{The two authors have contributed equally to this work.} \\
Beijing Institute of Genomics, \\
Chinese Academy of Sciences, 
Beijing, China, 100101}

\maketitle

\clearpage

\abstract
Cell growth in size is a complex process coordinated by intrinsic and environmental signals. In a recent work [Tzur et al., Science, 2009, 325:167-171], size distributions in an exponentially growing population of mammalian cells were used to infer the growth rate in size. The results suggest that cell growth is neither linear nor exponential, but subject to size-dependent regulation. To explain their data, we build a model in which the cell growth rate is controlled by the relative amount of mRNA and ribosomes in a cell. Plus a stochastic division rule, the evolutionary process of a population of cells can be simulated and the statistics of the \emph{in-silico} population agree well with the experimental data. To further explore the model space, alternative growth models and division rules are studied. This work may serve as a starting point for us to understand the rational behind cell growth and size regulation using predictive models.


\section*{Introduction}

Understanding the dynamical process of cell growth in size between divisions is a classic problem in biology. Over the decades there has been extensive research on this subject and yet much is still unknown about it~\cite{shields1978cell, sveiczer1996size, jorgensen2004cells, lloyd2013regulation, li2010mathematical}. Earlier attempts to measure growth rate at single-cell level suffer from technical limitations~\cite{killander1965quantitative, killander1965quantitativeb, conlon2003differences}. Now the state-of-art method can monitor cell size with much greater accuracy~\cite{son2012direct, mir2011optical}, but the intrinsic noise in cells and limited sample size that can be measured by experiment hindered the interpretation of the single-cell measurement data. Alternatively, the statistics of a population of synchronized or asynchronized cells can be accurately measured, from which the cell growth dynamics can be inferred~\cite{collins, conlon2003differences, kafri2013dynamics}. Together, these two types of 
measurements provide complementary data, shedding light on the mechanisms that regulate cell growth.

In a recent work, Tzur et~al.~\cite{Tzur} estimated the mean growth rate in size of a mouse lymphoblasts cell line (L1210) based on a population level approach. The rational behind this method is that, an asynchronous population growing exponentially (in number) has a steady size distribution, and the zero-flux condition of this steady distribution establishes a functional relation between the growth rate and size distributions of the asynchronous, newborn and dividing cell populations. This relation is known as the Collins-Richmond equation~\cite{collins} and has been used to estimate growth rate of bacteria and animal cells before~\cite{collins, koch1966distribution, anderson1969cell}. Compared with earlier work using similar approach, Tzur et~al. managed to remove all the unproven assumptions and obtain the cell size distributions in greater accuracy, thus significantly improve the fidelity of the results.

The estimated growth rate as a function of cell size from~\cite{Tzur} is replotted here in Fig.~\ref{fig_rate}A (Fig.~2A of the original paper). It appears that cells exhibit an exponential-like growth before their size reaching a certain threshold, after that the growth rate begin to drop (although a majority of cells have already divided before reaching this critical size). This ``$\Lambda$''-shaped growth pattern is consistent with previous results in~\cite{collins, anderson1969cell} and more recent results in~\cite{kafri2013dynamics} for different cell lines. Although the reduction of growth rate for large cells has been noticed before~\cite{collins, anderson1969cell}, a mathematical model that explicitly explain this growth pattern is still missing. In fact, previous work tend to treat this part of data as outlier, probably because (i) the data was not accurate enough for a quantitative analysis and (ii) only a small proportion of cells in the population are found in this rate reduction region (about 10\
as estimated in~\cite{anderson1969cell} and 35\% in~\cite{Tzur}). However, as we will show, albeit only affecting a small proportion of cells, this growth rate reduction can play an important role in maintaining cell size homeostasis. We suspect it might function as a regulatory mechanism for size control and is worth to be re-examined more carefully with the newly available experimental data.

In this paper we present a simple model to explain the experimental data in~\cite{Tzur}. The model assumes that the growth rate of a cell is determined by both ribosome number and mRNA level. Their relative abundance changes as cell-cycle progresses, coordinating the dynamics of cell growth. Plus an empirical division rule that tells a cell when to divide, the \emph{in-silico} cell population generated under our growth model can reproduce the observed experimental results, i.~e., the ``$\Lambda$''-shaped growth curve and cell size distributions of the asynchronous and newborn populations.

We emphasis that even though we build this model with some biological rational in our mind, it is still semi-phenomenological and serves mainly for the purpose of explaining the data. Mathematically, finding a model that can regenerate the observed data is an inverse problem and the solution (the model) is usually not unique. To this end, we also explore other phenomenological models as well as alternative division rules that may or may not give rise to results consistent with the experiment and explain why it is so.

\section*{A Simple Cell Growth Model}
\label{sec_simple}

\begin{figure}
 \centering
  \includegraphics[width=.9\textwidth]{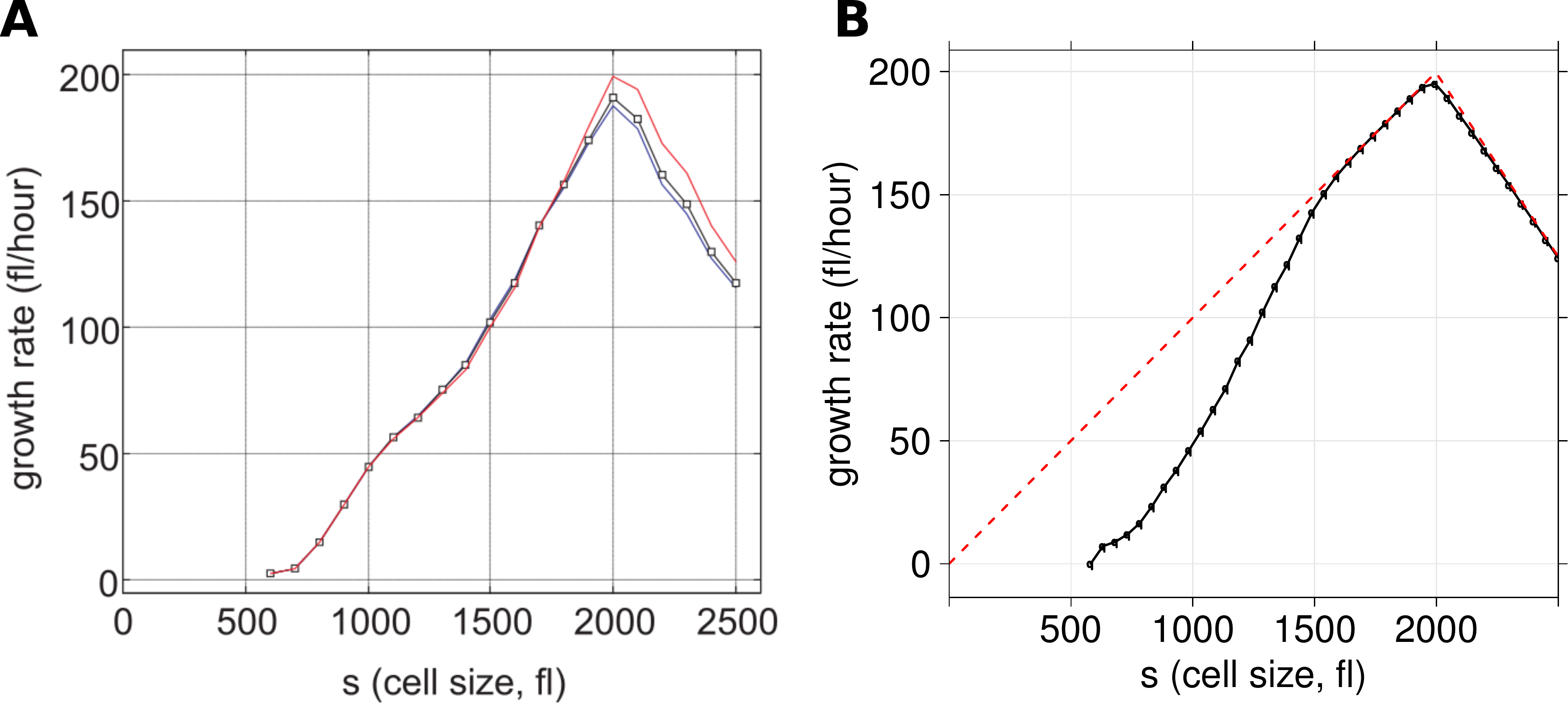}
 \caption{Cell growth rate as a function of cell size. 
 (A) Experimental result obtained using Collins-Richmond method in~\cite{Tzur} 
 (permission from AAAS to reuse this figure, different curves correspond
 to different detailed implementations).
 (B) Averaged growth rate obtained from the \emph{in-silico} population 
 simulated using our cell growth model. The dashed-red curve 
 represents pure exponential growth ($v(s) = \lambda_2 s$) for $0\le s < 2000$ and linear decay in growth rate ($v(s)=200-\gamma_2 s$) for $s\ge 2000$.}
 \label{fig_rate}
\end{figure}

Denote $v(s)$ as the cell growth rate as a function of cell size $s$ in an asynchronous population of cells in which the frequency density of any observable (size, age, growth rate, etc) is time-invariant. Since cells with the same size may have different growth rate, $v(s)$ should be understood as an ensemble average conditioning on the given size $\langle ds/dt \rangle_s$ (see Eq.~\eqref{eq_average_gr} for its mathematical definition). The estimated $v(s)$ from experimental measurements obtained by Tzur et~al.~\cite{Tzur} is shown in Fig.~\ref{fig_rate}A. It shows that $v(s)$ is neither a constant, which corresponding to linear growth, nor proportional to $s$, which corresponding to exponential growth. Instead, $v(s)$ is ``$\Lambda$''-shaped: it increases linearly with respect to $s$ when $s$ is small and then decreases after $s$ exceeds a certain threshold. We suspect this growth pattern is caused by some form of size-dependent regulation.

\begin{figure}
 \centering
 \includegraphics[width=.9\textwidth]{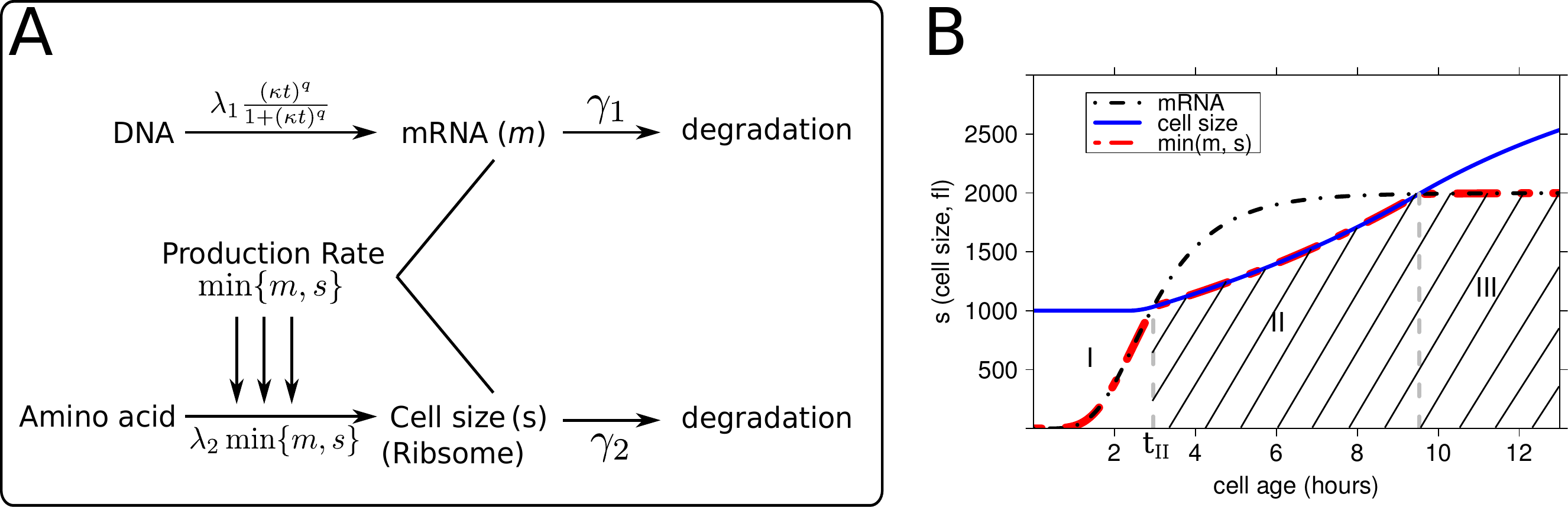}
 \caption{\textbf{(A)} A two-variable cell growth model. Cell size is proportional to the number of ribosomes it contains. The degradation rate per cell mass is $\gamma_2$ and the production rate is proportional to the number of working ribosomes, $\lambda_2 \min\{m, s\}$. \textbf{(B)} Trajectories of mRNA and cell size according to Eqs.~\eqref{eq_mRNA} and~\eqref{eq_ribosome}. Initially the mRNA level is set to zero. According to the relative abundance of mRNA and ribosomes, three growth stages can be identified in which mRNA and ribosomes play different roles in regulating cell growth (see main article).}
 \label{fig_ode}
\end{figure}

We propose a model sketched in Fig.~\ref{fig_ode}A. It is assumed that the size (volume) of a cell is proportional to its protein mass, and the later is further assumed to be proportional to the total number of ribosomes in the cell (thinking ribosome as a representative of proteome). Under the above assumptions, cell size, protein mass and ribosome number can be represented by one variable, $s$, after proper rescaling. The degradation rate per unit of protein mass is a constant $\gamma_2$. The protein synthesis rate is proportional to the total number of working ribosomes in the cell, i.~e., ribosomes that can allocate mRNA to initiate translation. The amount of mRNA is denoted by $m$, and its units is rescaled so that one unit of ribosome need one unit of mRNA. So the total working ribosomes in a cell equals to $\min\{m, s\}$ (for simplicity $m$ and $s$ are treated as continuous variable), and the protein synthesis rate is $\lambda_2 \min\{m, s\}$. The dynamics of mRNA is assumed to be age-dependent, with 
degradation rate $\gamma_1$ and production rate $\lambda_1 (\kappa t)^q/(1+ (\kappa t)^q)$. Here $t$ is the cell age, $q$ and $\kappa$ are two parameters. So we have
\begin{subequations}
\begin{align}
 \frac{dm}{dt} &= \frac{\lambda_1 (\kappa t)^q}{1+(\kappa t)^q} - \gamma_1 m, \label{eq_mRNA} \\
 \frac{ds}{dt} &= \left[\lambda_2 \min\{m, s\} - \gamma_2 s\right]^+, \label{eq_ribosome} 
\end{align}
\end{subequations}
where $[x]^+ = \max\{0, x\}$ keeps $ds/dt$ to be non-negative (even if protein degradation is faster than synthesis, the constituting amino acids remain in the cell, so the cell size will not shrink). The Hill's function term allows mRNA level to saturate quickly at a plateau. A typical solution of this system is shown in Fig.~\ref{fig_ode}B (see Methods for parameter values).

Initially mRNA level is low in the newborn (old mRNA degraded during mitosis and chromosomes need time to unfold). This is growth stage I in which insufficient mRNA supply limits protein synthesis. In stage II mRNA level builds up quickly, allowing all ribosomes to work full time, and the cell will grow exponentially. But the maximum mRNA level a cell can possibly support is limited. If the cell reaches a critical size without dividing, it will take up all the mRNA and its growth rate decreases in stage III.

To simulate an evolving population we also need to know how and when a cell divides. Following~\cite{Tzur} we assume the size difference between two sibling daughter cells obeys a Gaussian distribution $N(0, \sigma^2)$, with a standard deviation $\sigma \approx 68.8$ (fl) that is independent with the size of the mother cell. For convenience we set the mRNA level of the newborns to be 0, although a small non-zero value will give similar results. The detailed rules specifying when cell divides will be postponed in later sections as they have little effect in determining the shape of the growth rate curve. 

From the \emph{in-silico} cell population we obtain the mean growth rate $v(s)$ (Fig.~\ref{fig_rate}B. See Methods for detailed implementation). By tuning model parameters, quantitative agreement with the experiment result can be made.

\section*{Fitting the Growth Curve}
\label{sec_general}
Deducing a model that fits observed data is an inverse problem and the solution is never unique (as long as the model allows unlimited complexity). Nevertheless, giving the richness of information contained in our data, finding a simple model that fits all the data (growth rate and size distributions) is non-trivial. Next we analysis several models and explain why they can or cannot reproduce the growth curve.

In general, dynamics of a cell can be described by 
\begin{equation*}
 \frac{d\mathbf{X}}{dt} = \mathbf{A} (\mathbf{X}, t),
\end{equation*}
where the cell state $\mathbf{X}$ usually lives in a high dimension. $\mathbf{A}$ is some deterministic or stochastic operator. In an asynchronous population, the frequency density of $\mathbf{X}$, denoted by $p(\mathbf{x})$, is in steady state. Suppose in some experiment one can measure a component in $\mathbf{X}$, say $S$, while the rest components $\mathbf{M}$ are hidden variables. $S$ can be cell size, DNA mass, or surface marker intensity, etc. The Collins-Richmond equation can be applied on the frequency distribution of $S$ to get the growth rate of $S$, which is essentially in the following marginal form,
\begin{equation*}
 \bar{v}(s) \propto \frac{2 \bar{F}_0(s) - \bar{F}_{mi}(s) - \bar{F}_a(s)}{\bar{f}_a(s)}.
\end{equation*}
Here $\bar{f}_a(s), \bar{f}_{mi}(s)$ and $\bar{f}_0(s)$ (with $\bar{F}_a(s), \bar{F}_{mi}(s)$ and $\bar{F}_0(s)$ being their accumulative distributions) are the marginal of $p(\mathbf{x})$ (asynchronous population), $p_{mi}(\mathbf{x})$ (dividing cells) and $p_0(\mathbf{x})$ (newborns), respectively, i.~e.,
\begin{subequations}
\begin{align*}
 \bar{f}_a(s) &=  \int_\mathbf{m} p(s, \mathbf{m}) d\mathbf{m}, \\
 \bar{f}_{mi}(s) &=  \int_\mathbf{m} p_{mi}(s, \mathbf{m}) d\mathbf{m}, \\
 \bar{f}_0(s) &=  \int_\mathbf{m} p_0(s, \mathbf{m}) d\mathbf{m}.
\end{align*}
\end{subequations}
Intuitively $\bar{v}(s)$ is the average of growth rate of $S$, $dS/dt \equiv v(s, \mathbf{m})$, in the one-dimensional marginal space~\cite{kafri2013dynamics}. To put it in a more rigorous mathematical form, since the flux of $S$ across the point $S=s$ in the marginal space equals to the flux across the hyperplane $S=s$ in the full space where $\mathbf{X} = (S, \mathbf{M})$ lives, we have
\begin{equation*}
 \bar{v}(s) \bar{f}_a(s) = \int_\mathbf{m} v(s, \mathbf{m}) p(s, \mathbf{m}) d\mathbf{m}, \\ 
\end{equation*}
which gives 
\begin{equation}
 \bar{v}(s) = \frac{\int_\mathbf{m} v(s, \mathbf{m}) p(s, \mathbf{m}) d\mathbf{m}}{\bar{f}_a(s)} 
 = \int_\mathbf{m} v(s, \mathbf{m}) p(\mathbf{m}|s) d\mathbf{m} \equiv \left\langle \frac{ds}{dt} \right\rangle_s,
 \label{eq_average_gr}
\end{equation}
where $p(\mathbf{m}|s)$ is the conditional probability density of $\mathbf{M}$ given $S=s$. In other words, $\bar{v}(s)$ is the expectation of the growth rate $v(s, \mathbf{m})$ conditioning on a given $s$.

\begin{figure}
 \centering
 \includegraphics[width=.9\textwidth]{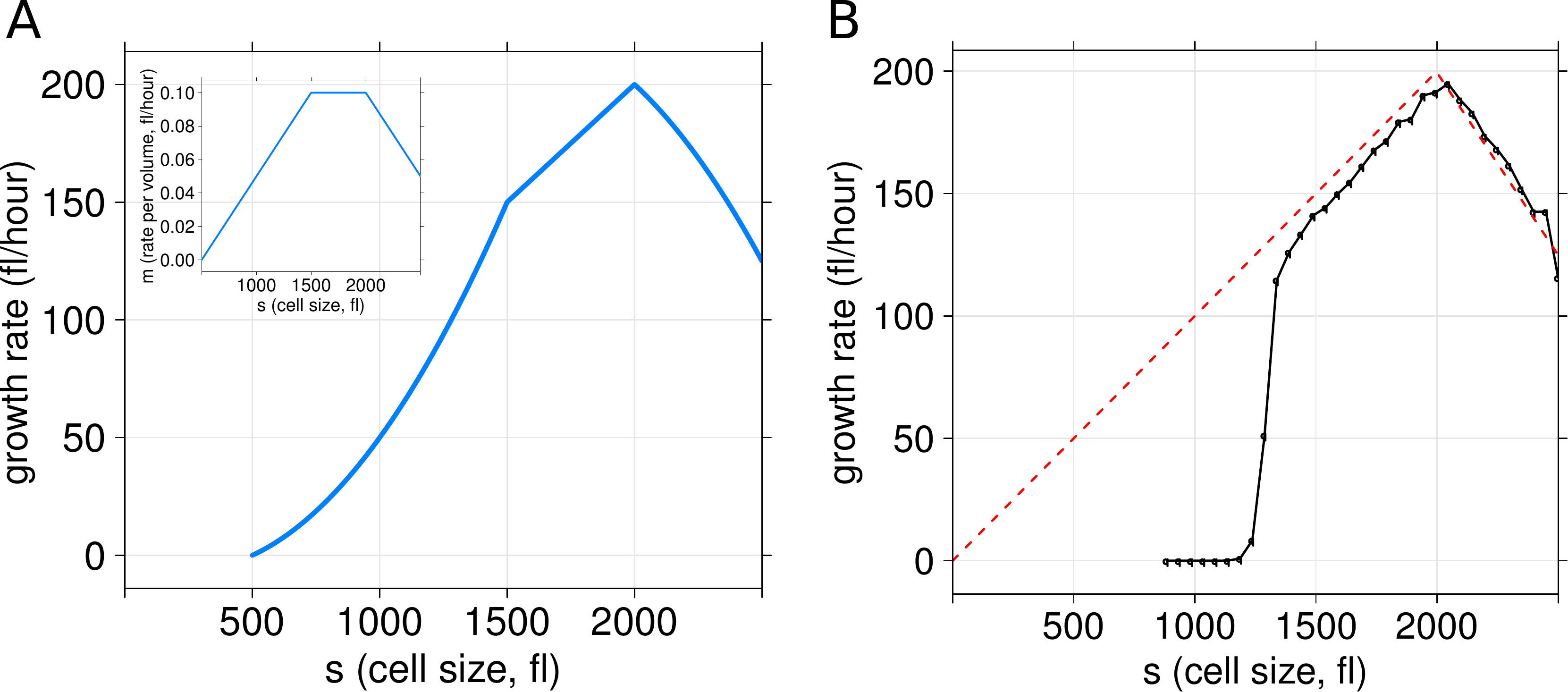}
 \caption{(A) Growth rate of model 3 as given by Eq.~\eqref{eq_rate_model3}.
 The insertion is the corresponding growth rate per volume ($m$ in Eq.~\eqref{eq_model3}).
 (B) Mean growth rate in model 4 (black curve).}
 \label{fig_model3}
\end{figure}

Next we consider a class of models with one hidden variable:
\begin{align*}
 \frac{dm}{dt} &= h(s,m, t), \\
 \frac{ds}{dt} &= v(s, m).
\end{align*}
Here $t$ denotes the cell age, $s$ is the cell size and the hidden variable $m$ may have different meaning in different models. The model given by Eqs.~\eqref{eq_mRNA} and \eqref{eq_ribosome}, which will be referred as \textbf{Model 1}, belongs to this class, with $m$ represents the mRNA level in a cell. Other examples includes:
\\
\\
\noindent \textbf{Model 2:}
\begin{equation}
 \frac{ds}{dt} = \lambda s - \gamma s, \label{eq_model2}
\end{equation}
where cell size $s$ is decoupled from any hidden variable.
\\
\\
\noindent \textbf{Model 3}
\begin{subequations}
\begin{align}
 m & = \begin{cases}
  \lambda + k_1(s - s_1), &\text{ if $s < s_1$}, \\ 
  \lambda, &\text{ if $s_1 \le s < s_2$}, \\
  \lambda + k_2(s - s_2), &\text{ if $s \ge s_2$}.
\end{cases} \label{eq_model3} \\
 \frac{ds}{dt} &= [ms]^+.
\end{align}
\end{subequations}
Here $m$ is the effective growth rate per unit of cell size and is chosen as a piecewise linear function of $s$ based on results from direct observation of single cell growth in~\cite{son2012direct}. For $\lambda = 0.1$, $k_1 = 0.0001$, $k_2 = -0.0001$, $s_1 = 1500$, $s_2 = 2000$, the $m\sim s$ relation is plotted in the insertion of Fig.~\ref{fig_model3}A. The shape of this curve is partially consistent with Fig. S4 in~\cite{son2012direct}, but we exaggerated over the part that $m$ decreases for large $s$, otherwise it cannot fit the data here. This inconsistence may be caused by different cell culture condition in the single-cell experiment compared with the population-level experiment (e.~g., loss of cell-cell interaction). Another possibility is that more than 65\% cells have already divided before reaching the critical size~\cite{Tzur}, so if the sample size in the single-cell measurement is small, this growth reduction may be missed.
\\
\\
\noindent \textbf{Model 4}
\begin{subequations}
\begin{align}
 m & \sim \mathcal{U}\left(-\frac{\lambda}{2}, \frac{\lambda}{2}\right) \text{, heritable}\\
 \frac{ds}{dt} &= 
 \begin{cases}
 0, &\text{ if $t < C - \ln 2/ (\lambda+m)$}, \\ 
 (\lambda + m)s, &\text{otherwise}.
 \end{cases}  \label{eq_rate_model4}
\end{align}
\label{eq_rate_model42}
\end{subequations}
It is briefly mentioned in~\cite{kafri2013dynamics} that the observed growth rate reduction for large cells might be caused by the factor that fast growing cells divide earlier at relative small size, leaving slow growing cells with relative large size behind. In the meanwhile all cells grow exponentially. To test this idea we build the above model in which cells have different intrinsic growth rate as indicated by $m$, a random variable uniformly distributed in $(-\lambda/2, \lambda/2), \lambda=0.1$. When dividing, the same $m$ is passed from a mother cell to the daughter cells. We have to use a special division rule (see Methods) to make the fastest growing cells divide at $s=2000$ and the slowest-growing cells divide at $s=2500$ so as to fit the growth curve. One problem caused by different growth rate in this model is that the time needed for a cell to double its size via exponential growth is different (which is $\ln 2 / (\lambda + m)$). To give no selective bias in the heterogeneous population, we add 
an idle time with length $C - \ln 2/ (\
lambda+m), C=14$ (hour), to each cell before it starts to grow, so that cells would have equal cell cycle length.
\\
\\
For each model, we check if the model-predicted growth rate 
\begin{equation}
 \bar{v}(s) = \int_m v(s, m) p(m|s)dm, \label{eq_average_gr6}
\end{equation}
agrees with the experimental result (Fig.~\ref{fig_rate}A). For model 2, replacing $v(s, m)$ in the above equation by Eq.~\eqref{eq_model2} immediately gives $\bar{v}(s) = (\lambda - \gamma)s$. It gives a straight line so is inconsistent with the experiment.

For model 3, since $m$ depends on $s$ only,
$p(m|s) = \delta (m - m(s))$, and Eq.~\eqref{eq_average_gr6} can be
computed explicitly as
\begin{equation}
 \bar{v}(s) = \begin{cases}
  [\lambda_1 s + k_1(s - s_1)s]^+, &\text{ if $s < s_1$}, \\ 
  \lambda_1 s , &\text{ if $s_1 \le s < s_2$}, \\
  \lambda_1 s + k_2(s - s_2)s, &\text{ if $s \ge s_2$}.
\end{cases} 
\label{eq_rate_model3}
\end{equation}
This growth curve agrees with the experiment 
reasonably well (Fig.~\ref{fig_model3}A).

For model 4 computer simulation shows that the averaged growth rate $\bar{v}(s)$ is indeed ``$\Lambda$''-shaped (Fig.~\ref{fig_model3}B). In fact, within the range $1300 < s < 2000$, most cells are in exponential growth stage and no cell divide, so $m$ is uniformly distributed and $\bar{v}(s) = \lambda s$. As $s$ increases, the mean value of $m$ in the population shifts from 0 to $-\lambda/2$ because cells with larger $m$ start to divide, so the mean growth rate decreases. However, inconsistent with the experiment, for $s$ smaller than 1300 the mean growth rate is close to zero because many newborns are idle. In addition, the cell size distributions under this model differ from the measured distributions (result not shown). Thus the model in its current form only provides partial explanation to the experiment.

For model 1, explicit expression for $\bar{v}(s)$ is not available, but some qualitative analysis can still be made. On one hand, for small $s$ ($500 < s < 1500$), replace the $v(s,m)$ in Eq.~\eqref{eq_average_gr6} with  Eq.~\eqref{eq_ribosome} and split the integral, we have
\begin{equation*}
  \bar{v}(s)  = \int_0^s (\lambda_2 m - \gamma_2 s)^+ p(m|s) dm + \int_s^{m_{max}} (\lambda_2 - \gamma_2) s p(m|s) dm,
\end{equation*}
where $m_{max} \approx 2000$ is the maximum mRNA level. For small cells ($s<1500$) the mRNA level tend to be low (if they are in growth stage I), so $p(m|s) > 0$ and $(\lambda_2 m - \gamma_2 s)^+ < (\lambda_2 - \gamma_2)s$ for $0<m<s$. Overall $\bar{v}(s) < (\lambda_2 - \gamma_2)s$, which explains why for small $s$ the growth curve (black-solid line in Fig.~\ref{fig_rate}B) lies below the exponential curve (red-dashed line in Fig.~\ref{fig_rate}B). On the other hand, for large cells ($s \ge 1500$) the mRNA level tend to be saturated around $m_{max}$, approximately we have $p(m|s) \approx \delta(m-m_{max})$ and $\bar{v}(s) = \lambda_2 \min(s, m_{max}) - \gamma_2 s$.
So for $1500 \le s < 2000$ $\bar{v}(s)$ increases linearly with $s$ (an exponential growth with rate $\lambda_2 - \gamma_2$) and for $s \ge 2000$ $\bar{v}(s)$ decreases linearly with $s$ (with a slope equals to $\gamma_2$). This is also the way how the parameters $\lambda_1, \gamma_1$, which controls $m_{max}$, and $\lambda_2, \gamma_2$ are chosen by comparing with the experimental growth curve in Fig.~\ref{fig_rate}A.

\section*{Division Rules and Size Homeostasis}
\label{sec_div}
A division rule tells a cell when to divide. It effects the size homeostasis in a cell population. Here we ask, for cell growth model 1, what kind of division rule can reproduce the asynchronous and newborn size distributions that have been directly measured by experiment in~\cite{Tzur}.

We assume there exists a division rate function $p(\mathbf{x}, t)$ that depends on cell state $\mathbf{x}$ and cell age $t$. The probability that a cell divides during an infinitesimal time interval $dt$ is $p(\mathbf{x}, t)dt$. In particular, we consider the following division rules.
\begin{enumerate}
 \item \textbf{Age-gate:} \\
 $$
 p(t) = \begin{cases}
  0, &\text{ if $t < t_0$}, \\ 
  p_0 , &\text{ if $t \ge t_0$}.
\end{cases} 
 $$
 \item \textbf{Age-gate plus size-gate:} \\
 $$p(s, t) = p_1(t) + p_2(s),$$
 with
 $$
 p_1(t) = \begin{cases}
  0, &\text{ if $t < t_0$}, \\ 
  p_0 , &\text{ if $t \ge t_0$},
\end{cases} 
\ \ \ \ \ \ \ \ \ \ 
 p_2(s) = \begin{cases}
  0, &\text{ if $s < s_0$}, \\ 
  p_0 , &\text{ if $s \ge s_0$}.
\end{cases} 
 $$
 \item \textbf{Signal integration:} \\
 $$
 p(t) =  \begin{cases}
  0, &\text{ if $A(t) < A_0$}, \\ 
  p_0, &\text{ if $A(t) \ge A_0$},
\end{cases}
 $$
where
 $$
 A(t) = \int_{t_{II}}^t \min\{ m, s\} dt'
 $$ 
 is the area of the part of the shaded region in Fig.~\ref{fig_ode}B up to $t$.
 \end{enumerate}

\begin{figure}
 \centering
 \includegraphics[width=.9\textwidth]{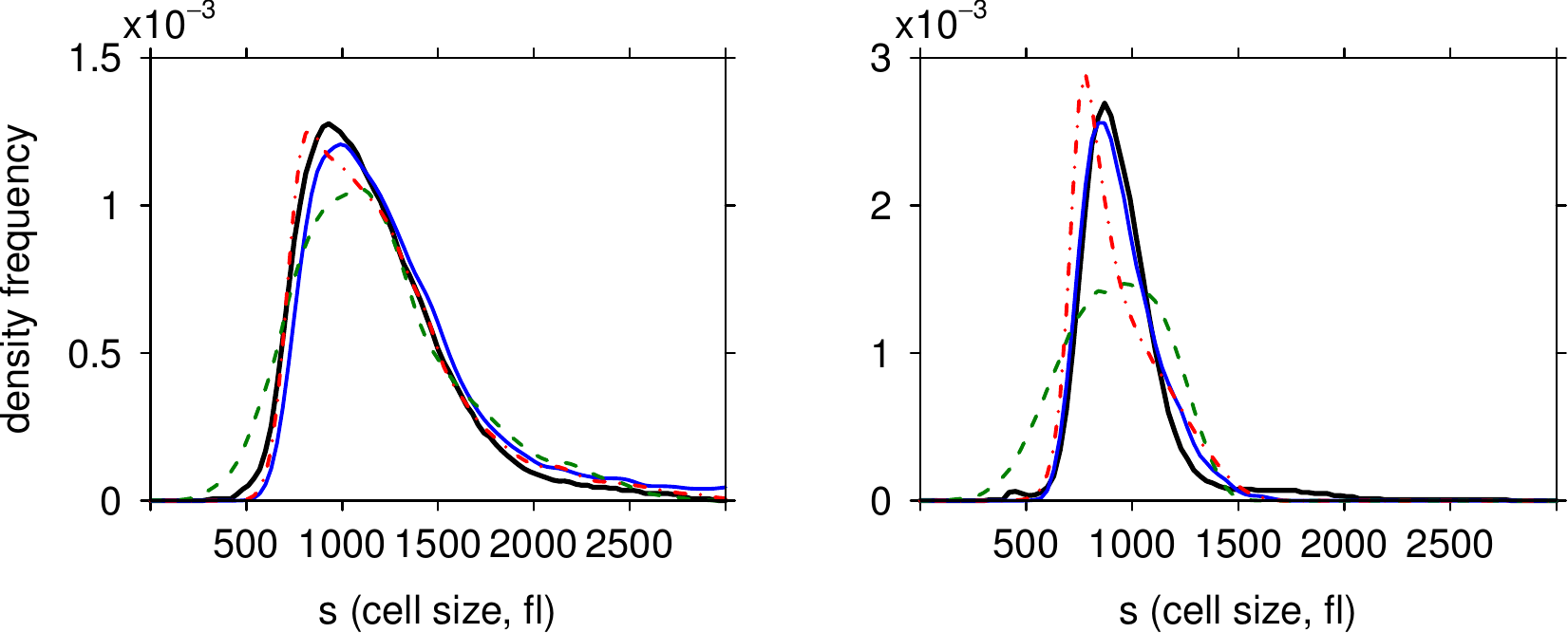}
 \caption{Asynchronous (left) and newborn (right) cell size distributions from direct measurement (black thick line) and \emph{in-silico} population simulated using division rule 1 (green dashed line), rule 2 (red dot-dashed line) and
 rule 3 (blue solid line). Sample size is $N=10^5$. See Methods for the parameter values.}
 \label{fig_dists}
\end{figure}

For each division rule, we search for the parameters ($p_0$ and $t_0$ for rule 1, $p_0, s_0$ and $t_0$ for rule 2, $p_0$ and $A_0$ for rule 3) that minimize the $L_1$-distance between the \emph{in-slico} and experimental distributions. The best-fit results are shown in Fig.~\ref{fig_dists}. See Methods for the optimization procedure.

Under division rule 1, a homeostasis population can be established. Interestingly, if we apply the same division rule to an exponential growth population (model 2), there is no stable size distribution. This is consistent with the result in~\cite{anderson1969cell} saying that, if cells grow exponentially in size and the division rule depends on age only, the variance of cell size in a population will diverge. In this sense the ``$\Lambda$''-shaped growth pattern can be understood as a regulation mechanism bounding the cell size from above, which leads to size homeostasis even for a division rule that depends only on cell age. However, by itself this mechanism is not good enough as the \emph{in-silico} size distributions (Fig.~\ref{fig_dists} green dashed curves) cannot match the measured ones very well. In particular, the size distribution of the newborns is much wider and some newborns are relatively smaller in size. These cells born to be too small are likely to find themselves in a disadvantageous place 
to start with compared with other cells. In other words, the quality of the population produced by this 
division rule is compromised. The reason is that there is no quality-checking (in terms of cell size) in this scheme.

Under division rule 2, the \emph{in-silico} distributions agree with the experimental data reasonably well (Fig.~\ref{fig_dists} red dot-dashed curves). It degenerates to rule 1 when $s_0 \rightarrow \infty$. So by taking extra size information into account, this division scheme achieves better agreement with experimental data than the one that only uses age information.

Division rule 3 also leads to a good fitting (Fig.~\ref{fig_dists} blue solid curves). It is assumed in this scheme that after a cell leaves region I (Fig.~\ref{fig_ode}B), it begins to measure a mitosis-signal in an integral way. Here we take the signal to be proportional to the protein synthesis rate, $\min\{s, m\}$. Once the time integral of this signal reaches a critical value, the cell begins to divide with a constant rate $p_0$. (If the area within region I is also taken into the integration, similar results still hold.) 

\begin{figure}
 \centering
 \includegraphics[width=.6\textwidth]{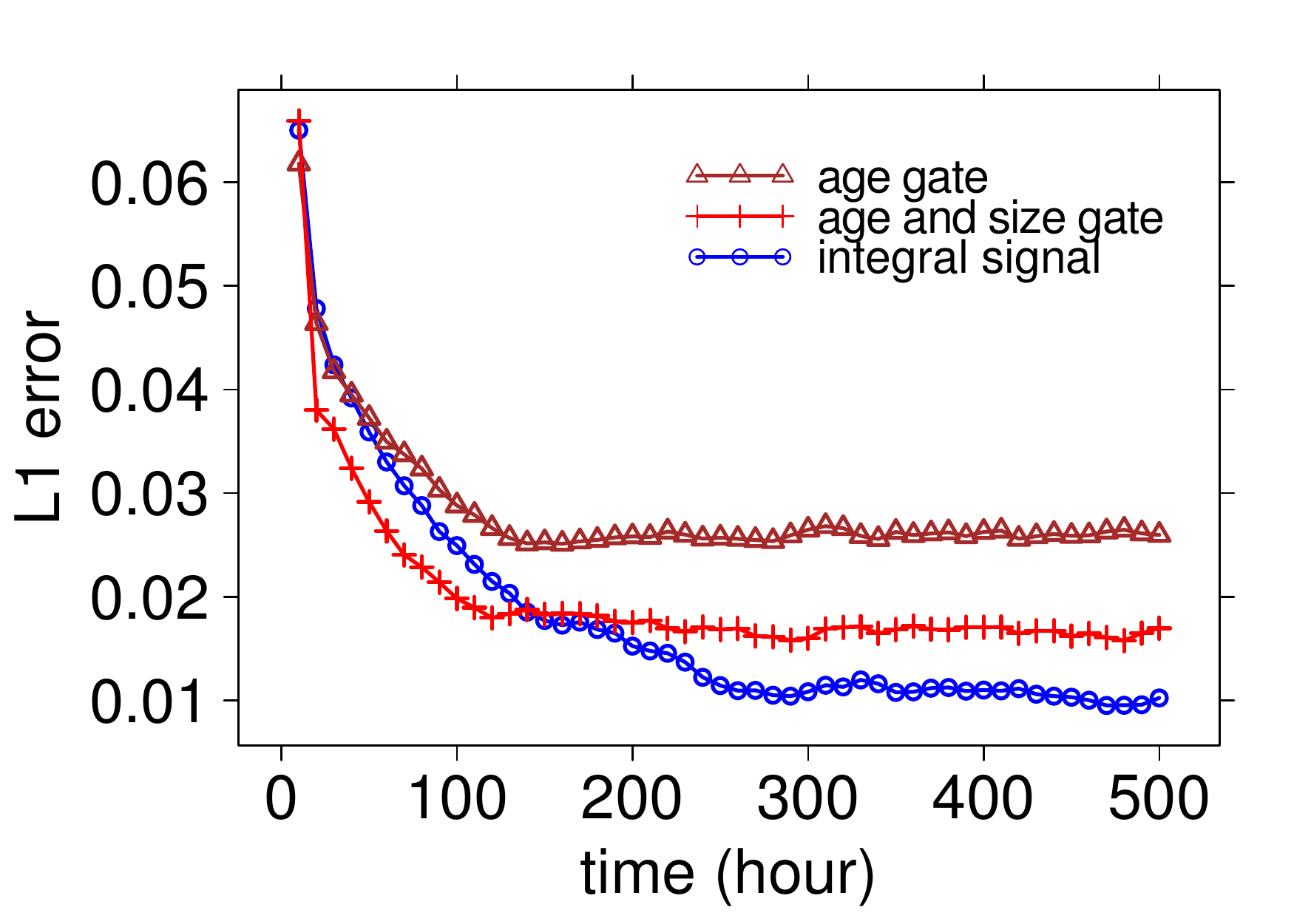}
 \caption{$L_1$-distance between the \emph{in-silico} and experimental size distributions (see Methods). Initially the population is synchronized at age zero and all cells have identical cell size. Different curves correspond to different division rules.}
 \label{fig_conv}
\end{figure}

For each division rule (using the optimal parameters we found), the $L_1$-distance between the \emph{in-silico} and experiment distributions as a function of simulation time is plotted in Fig.~\ref{fig_conv}. Initially all cells are identical in size and synchronized at age zero. As the population evolves under the growth model and division rule, the size distributions gradually reach homeostasis. It appears division rule 3 gives the best fit to the experimental data. For all division rules, the time taken for the \emph{in-silico} population to reach size homeostasis is around 10 days. Given that the average cell cycle length is roughly 10 hours in our model (see Fig.~\ref{fig_rate}B), more than 20 rounds of divisions are needed for a synchronous population to reach homeostasis.

Overall, for growth model 1, both division rules 2 and 3 explain the experimental data reasonably well and they both predict that, for cells with the same size, older cells are more likely to divide than younger cells, and for cells with the same age, larger cells are more likely to divide than smaller cells, which is consistent with the observation made in~\cite{Tzur}. Again, we note that there could be many other division rules which can match the data.

\section*{Discussions}
In this work we proposed a simple cell growth model aiming to explain the ``$\Lambda$''-shaped growth rate curve observed in experiments. It makes an excellent demonstration that complex experimental results can sometimes be captured by very simple intuition. In our model the growth rate of a cell is regulated by its mRNA content: when there is enough mRNA, cell enjoys an exponential growth, otherwise its growth rate is compromised. The rate-limiting effect of mRNA occurs when a cell is newly born or its size exceeds a certain threshold. The statistics of the \emph{in-silico} population generated using our growth model matches with the experimental results very well.

While our motivation in building the model is to explain the data, two central features of the model may resemble the actual biological mechanisms regulating cell growth. The first one is a transit slow-growing period right after a cell is born, a behavior which is observed in several cell lines~\cite{kafri2013dynamics}. There could be many explanations for this and that the mRNA is playing a rate-limiting role is one of them. After all, during mitosis the production of mRNA should be minimum as the chromosomes are tightly packed. In addition, chromosome-unfolding and transcription initialization also delay the mRNA production in the newborns. The other crucial feature of the model is the existence of a maximum mRNA level that a cell can reach.  Recent experiments show that the mRNA output reaches a plateau in division arrested yeast because of limited DNA copy number~\cite{zhurinsky2010coordinated}. Together, the interplay between mRNA and protein synthesis provide a 
potential mechanism for growth control, consistent with the idea that gene expression and cell size are tightly correlated~\cite{marguerat2012coordinating}.

Cell size homeostasis is coordinated by growth and division. We studied several empirical division rules to see if they can regenerate the experimental data under our growth model. It turns out that the observed size distributions cannot be explained simply by an age-gate division rule and certain quality control mechanism, or ``sizer'', in mitosis-decision is needed. The division rule based on the time integral of a growth signal fits the experiment very well. Interestingly, there is evidence showing budding yeast takes a similar approach in its division control~\cite{tangchao}.

%

The Collins-Richmond relation provides an elegant way of extracting dynamic properties of some observable from its frequency distribution in an asynchronous population. Examples of data on which this method could be applied include DNA content~\cite{dean1974mathematical} and RNA Polymerase II distribution from ChIP-seq experiments~\cite{ehrensberger2013mechanistic}. We described a general framework for consistency checking between a growth model and results based on the Collins-Richmond method. It can be useful in providing some heuristic insight or guidance in building and tuning more complex models. 

Finally we note that, the biggest limitation of the Collins-Richmond method in its current form is that it can only handle one variable at a time and is thus unable to directly capture the dynamical interaction between different variables. Recently Kafri et~al.~\cite{kafri2013dynamics} developed a new framework called ERA (ergodic rate analysis) that extends the density balance law, based on which the Collins-Richmond equation holds, to high-dimensional data. They were able to apply this new method on asynchronous population of Hela cells which revealed more insights on cell cycle 
dynamics.

\section*{Methods}
\subsection*{Simulation of cell population}
We want to simulate the evolution of a population of cells and collect its statistical information when cell size reaches homeostasis. It is practically impossible to simulate the entire population because the number of cells grows exponentially. Instead, we keep track of only a limited number of cells. In particular, we maintain a population of $N=10^5$ cells, and randomly delete one cell among them with equal chance whenever there is a cell division. Essentially this is the Moran population process~\cite{moran1958random} with fixed population size. It mimics drawing random samples from an exponentially growing population provided that there is no inheritable fitness difference between cell lineages, so the statistics thus obtained will be the same as that of the total population (apart from a sampling error).

To simulate the growth of each cell, Eqs.~\eqref{eq_mRNA} and \eqref{eq_ribosome} is solved using fourth order Rugger-Kutta method with constant step-size $\Delta t =0.05 $ (hour). The parameters used are: $\lambda_1=2000$, $\gamma_1=1$, $\lambda_2=0.25$, $\gamma_2=0.15$, $\kappa=0.5$, $q=4$. These values are chosen empirically to fit the experiment. In principle the number of mRNA molecules and ribosomes in a cell are integers and driven by stochastic processes, which can be described by a stochastic chemical reaction system. However, since their copy numbers are huge in a cell, a continuous and deterministic approach makes a very good approximation here. Additionally, since we are mainly interested about the statistics of the population, the random fluctuations in the individual cells are averaged out and have almost no effect on the growth rate and size distributions.

While the growth of individual cells can be approximated by deterministic ordinary differential equations, cell divisions need to be treated as stochastic events that arrive random in time. From the division rule we can compute the division rate $p(\mathbf{x}, t)$ for a cell with state $\mathbf{x}$ at age $t$, and the probability that the cell divides during a small time interval $dt$ is $p(\mathbf{x}, t) dt$. So a straight-forward but less efficient implementation is to generate a random variable $u$ uniformly distributed in $(0,1)$, and compare it with the mitosis probability $p(\mathbf{x}, t)dt$. Only if $u<p(\mathbf{x}, t) dt$ then the cell divides. Doing so we need to generate one random variable for each cell at every time step thus is very computational expensive for large population. A more efficient way is to assign an uniformly distributed random variable $u$ to each newborn cell and monitor the value
$$
\int_0^{\tau} p(\mathbf{x}, t) dt + \ln u,
$$
for this cell. This term is negative when $\tau$ is small and increases as $\tau$ marching forward. At the time when its value reaches zero the cell divides. It can be shown that the two procedures generate statistically equivalent waiting time for mitosis as time step-size approaches zero~\cite{haseltine2002approximate}. Now for one cell we only need to generate one random variable in its whole cell cycle thus the simulation speed is much faster.

\subsection*{Collecting statistics from the \emph{in-silico} population}
\textbf{Growth rate curve:} For each cell we record its size $s(t)$ at each time-step $t_n$. The growth rate of an individual cell at age $t_n$ is estimated by $(s(t_{n}) - s(t_{n-1}))/\Delta t$. To compute the average growth rate $v(s)$ as a function of cell size $s$, we sort cells by size, partition them into small intervals of $s$, and average the growth rate of cells within each interval.

\noindent\textbf{Asynchronous and newborn size distributions:}  The asynchronous size distribution can be directly sampled from the homeostasis \emph{in-silico} population. For the size distribution of the newborns, we collect and record the size information of the newborns during the simulation, and when needed, the most recent $N=10^5$ newborns in the database are used to sample the size distribution.

\subsection*{Fitting the parameters in division rules}
Each division rule contains several parameters. For a given set of parameters of a division rule, we simulate the population and obtain its asynchronous size distribution $f^{sim}_a$ and newborn distribution $f^{sim}_0$ as described above. Define the error as the sum of $L_1$-distance between $f^{sim}_{a,0}$ and the experimental results $f^{exp}_{a,0}$,
$$
Err \equiv \left\| f^{sim}_a - f^{exp}_a \right\|_1 + \left\| f^{sim}_0 - f^{exp}_0 \right\|_1.
$$
Minimizing the error within the parameter space is a nonlinear optimization problem and we use the ``optim'' routine in R for this task. The parameters we found are: division rule 1, $t_0 = 6.4, p_0=0.8$; division rule 2, $t_0 = 8, s_0 = 1440, p_0=0.18$, division rule 3, $A_0 = 6400, p_0=0.5$.

We also tested the $L_\infty$, $L_2$ and Kullback-Leibler distances. The first two have similar result as the $L_1$-distance, but the last one seems to be a little less sensitive for our case.

\subsection*{Division rule for model 4}
 $$
 p(t) =  \begin{cases}
  0, &\text{ if $s(\lambda + m+ 0.35) \le 1000$}, \\ 
  p_0, &\text{ if $s(\lambda + m+0.35) > 1000$}.
\end{cases}
  $$
  $\lambda=0.1$, $m\sim \mathcal{U}(-0.05, 0.05)$ 
  and $p_0=10$ (see Eq.\eqref{eq_rate_model42}).
  It is easy to verify that, the fastest growing cell ($m=0.05$) divides around size $s=2000$, and the slowest growing cell ($m=-0.05$) divides
  around size $s=2500$.

\section*{Acknowledgment}
YH are thankful to A. Lander for introducing this topic and to S. Wang for her initiative programing work. We thank R. Kafri, J. Lei, C. Tang and X. Wang for helpful discussions. We also thank the anonymous reviewers for their helpful comments. YH is supported by the NSFC (National Science Foundation of China) under Grant No. 11301294 and TZ is supported by NSFC under Grant No. 31301093.
\bibliography{ref}

\begin{thebibliography}{22}
\providecommand{\url}[1]{\texttt{#1}}
\providecommand{\urlprefix}{ }

\bibitem[Shields et~al.(1978)Shields, Brooks, Riddle, Capellaro, and
  Delia]{shields1978cell}
Shields, R., R.~F. Brooks, P.~N. Riddle, D.~F. Capellaro, and D.~Delia, 1978.
\newblock Cell size, cell cycle and transition probability in mouse
  fibroblasts.
\newblock \emph{Cell} 15:469--474.

\bibitem[Sveiczer et~al.(1996)Sveiczer, Novak, and Mitchison]{sveiczer1996size}
Sveiczer, A., B.~Novak, and J.~Mitchison, 1996.
\newblock The size control of fission yeast revisited.
\newblock \emph{Journal of Cell Science} 109:2947--2957.

\bibitem[Jorgensen and Tyers(2004)]{jorgensen2004cells}
Jorgensen, P., and M.~Tyers, 2004.
\newblock How cells coordinate growth and division.
\newblock \emph{Current Biology} 14:R1014--R1027.

\bibitem[Lloyd(2013)]{lloyd2013regulation}
Lloyd, A.~C., 2013.
\newblock The Regulation of Cell Size.
\newblock \emph{Cell} 154:1194--1205.

\bibitem[Li et~al.(2010)Li, Shao, Yu, Ouyang, and Wang]{li2010mathematical}
Li, B., B.~Shao, C.~Yu, Q.~Ouyang, and H.~Wang, 2010.
\newblock A mathematical model for cell size control in fission yeast.
\newblock \emph{Journal of Theoretical Biology} 264:771--781.

\bibitem[Killander and
  Zetterberg(1965{\natexlab{a}})]{killander1965quantitative}
Killander, D., and A.~Zetterberg, 1965.
\newblock Quantitative cytochemical studies on interphase growth: I.
  Determination of DNA, RNA and mass content of age determined mouse
  fibroblasts in vitro and of intercellular variation in generation time.
\newblock \emph{Experimental cell research} 38:272--284.

\bibitem[Killander and
  Zetterberg(1965{\natexlab{b}})]{killander1965quantitativeb}
Killander, D., and A.~Zetterberg, 1965.
\newblock A quantitative cytochemical investigation of the relationship between
  cell mass and initiation of DNA synthesis in mouse fibroblasts in vitro.
\newblock \emph{Experimental cell research} 40:12--20.

\bibitem[Conlon and Raff(2003)]{conlon2003differences}
Conlon, I., and M.~Raff, 2003.
\newblock Differences in the way a mammalian cell and yeast cells coordinate
  cell growth and cell-cycle progression.
\newblock \emph{Journal of biology} 2:7.

\bibitem[Son et~al.(2012)Son, Tzur, Weng, Jorgensen, Kim, Kirschner, and
  Manalis]{son2012direct}
Son, S., A.~Tzur, Y.~Weng, P.~Jorgensen, J.~Kim, M.~W. Kirschner, and S.~R.
  Manalis, 2012.
\newblock Direct observation of mammalian cell growth and size regulation.
\newblock \emph{Nature Methods} 9:910--912.

\bibitem[Mir et~al.(2011)Mir, Wang, Shen, Bednarz, Bashir, Golding, Prasanth,
  and Popescu]{mir2011optical}
Mir, M., Z.~Wang, Z.~Shen, M.~Bednarz, R.~Bashir, I.~Golding, S.~G. Prasanth,
  and G.~Popescu, 2011.
\newblock Optical measurement of cycle-dependent cell growth.
\newblock \emph{Proceedings of the National Academy of Sciences}
  108:13124--13129.

\bibitem[Collins and Richmond(1962)]{collins}
Collins, J., and M.~Richmond, 1962.
\newblock Rate of growth of Bacillus cereus between divisions.
\newblock \emph{Journal of general microbiology} 28:15--33.

\bibitem[Kafri et~al.(2013)Kafri, Levy, Ginzberg, Oh, Lahav, and
  Kirschner]{kafri2013dynamics}
Kafri, R., J.~Levy, M.~B. Ginzberg, S.~Oh, G.~Lahav, and M.~W. Kirschner, 2013.
\newblock Dynamics extracted from fixed cells reveal feedback linking cell
  growth to cell cycle.
\newblock \emph{Nature} 494:480--483.

\bibitem[Tzur et~al.(2009)Tzur, Kafri, LeBleu, Lahav, and Kirschner]{Tzur}
Tzur, A., R.~Kafri, V.~S. LeBleu, G.~Lahav, and M.~W. Kirschner, 2009.
\newblock Cell growth and size homeostasis in proliferating animal cells.
\newblock \emph{Science} 325:167--171.

\bibitem[Koch(1966)]{koch1966distribution}
Koch, A., 1966.
\newblock Distribution of cell size in growing cultures of bacteria and the
  applicability of the Collins-Richmond principle.
\newblock \emph{Journal of General Microbiology} 45:409--417.

\bibitem[Anderson et~al.(1969)Anderson, Bell, Petersen, and
  Tobey]{anderson1969cell}
Anderson, E., G.~Bell, D.~Petersen, and R.~Tobey, 1969.
\newblock Cell growth and division: {IV}. {D}etermination of volume growth rate
  and division probability.
\newblock \emph{Biophysical journal} 9:246--263.

\bibitem[Zhurinsky et~al.(2010)Zhurinsky, Leonhard, Watt, Marguerat,
  B{\"a}hler, and Nurse]{zhurinsky2010coordinated}
Zhurinsky, J., K.~Leonhard, S.~Watt, S.~Marguerat, J.~B{\"a}hler, and P.~Nurse,
  2010.
\newblock A coordinated global control over cellular transcription.
\newblock \emph{Current Biology} 20.

\bibitem[Marguerat and B{\"a}hler(2012)]{marguerat2012coordinating}
Marguerat, S., and J.~B{\"a}hler, 2012.
\newblock Coordinating genome expression with cell size.
\newblock \emph{Trends in Genetics} 28:560--565.

\bibitem[Liu et~al.()Liu, Wang, Yang, Liu, Qu, Hu, Ouyang, and Tang]{tangchao}
Liu, X., X.~Wang, X.~Yang, S.~Liu, Y.~Qu, L.~Hu, Q.~Ouyang, and C.~Tang.
\newblock Reliable cell cycle commitment in budding yeast is ensured by signal
  integration.
\newblock \emph{submitted} .

\bibitem[Dean and Jett(1974)]{dean1974mathematical}
Dean, P.~N., and J.~H. Jett, 1974.
\newblock Mathematical analysis of DNA distributions derived from flow
  microfluorometry.
\newblock \emph{The Journal of cell biology} 60:523--527.

\bibitem[Ehrensberger et~al.(2013)Ehrensberger, Kelly, and
  Svejstrup]{ehrensberger2013mechanistic}
Ehrensberger, A.~H., G.~P. Kelly, and J.~Q. Svejstrup, 2013.
\newblock Mechanistic Interpretation of Promoter-Proximal Peaks and RNAPII
  Density Maps.
\newblock \emph{Cell} 154:713--715.

\bibitem[Moran(1958)]{moran1958random}
Moran, P. A.~P., 1958.
\newblock Random processes in genetics.
\newblock \emph{In} Mathematical Proceedings of the Cambridge Philosophical
  Society. Cambridge Univ Press, volume~54, 60--71.

\bibitem[Haseltine and Rawlings(2002)]{haseltine2002approximate}
Haseltine, E.~L., and J.~B. Rawlings, 2002.
\newblock Approximate simulation of coupled fast and slow reactions for
  stochastic chemical kinetics.
\newblock \emph{The Journal of chemical physics} 117:6959.

\end{thebibliography}
\end{document}